\documentclass[aps,a4paper, preprint, superscriptaddress,preprintnumbers,floatfix,nofootinbib,amsmath,amssymb]{revtex4-1}

\usepackage{url}
\usepackage{hyperref}
\usepackage{color}
\usepackage{cancel}
\usepackage{cleveref}
\usepackage{subfig}
\usepackage{soul}
\usepackage[normalem]{ulem}

\usepackage{amstext,amssymb}
\usepackage{amsmath}
\usepackage{graphicx}
\usepackage{xspace}
\usepackage{color}
\usepackage{units}
\usepackage[T1]{fontenc}
\usepackage{amsmath,bm}
\usepackage{nicefrac}
\usepackage{slashed} 
\usepackage{multirow}
\newcommand{\be}{\begin{equation}}
\newcommand{\ee}{\end{equation}}
\newcommand{\bea}{\begin{eqnarray}}
\newcommand{\eea}{\end{eqnarray}}
\newcommand{\nn}{\nonumber}

\def\s1{\hat s}

\usepackage{slashed}

\newcommand{\nua}[1]{\ensuremath{\rlap{\kern-2.5pt\ensuremath{\overset{\scriptscriptstyle(-)}{\phantom{\nu}}}}{\ensuremath{{\nu}_{#1}}}}\xspace}

\definecolor{brickred}{rgb}{0.8, 0.25, 0.33}
\definecolor{brightcerulean}{rgb}{0.11, 0.67, 0.84}
\definecolor{brown(traditional)}{rgb}{0.59, 0.29, 0.0}
\begin{document}
\title{Exploring Models with Modular Symmetry in Neutrino Oscillation Experiments}
\author{Priya Mishra }
\email{mishpriya99@gmail.com}
\affiliation{School of Physics, University of Hyderabad, Hyderabad - 500046, India}
\author{Mitesh Kumar Behera}
\email{miteshbehera1304@gmail.com}
\affiliation{Department of Physics, Faculty of Science, Chulalongkorn University, Bangkok-10330, Thailand}
\author{Papia Panda}
\email{ppapia93@gmail.com}
\affiliation{School of Physics,  University of Hyderabad, Hyderabad - 500046,  India}
\author{Monojit Ghosh}
\email{mghosh@irb.hr}
\affiliation{Center of Excellence for Advanced Materials and Sensing Devices, Ru{\dj}er Bo\v{s}kovi\'c Institute, 10000 Zagreb, Croatia}
\author{ Rukmani Mohanta}
\email{rmsp@uohyd.ac.in}
\affiliation{School of Physics,  University of Hyderabad, Hyderabad - 500046,  India}

\begin{abstract}
 Our study aims to investigate the viability of neutrino mass models that arise from discrete non-Abelian modular symmetry groups, i.e., $\Gamma_N$ with ($N=1,2,3,\dots$) in the future neutrino experiments T2HK, DUNE, and JUNO. Modular symmetry reduces the usage of flavon fields compared to the conventional discrete flavor symmetry models. Theories based on modular symmetries predict the values of leptonic mixing parameters, and therefore, these models can be tested in future neutrino experiments. In this study, we consider three models based on the $A_4$ modular symmetry, i.e., Model-A, B, and C such a way that they predict different values of the oscillation parameters but still allowed with respect to the current data. In the future, it is expected that T2HK, DUNE, and JUNO will measure the neutrino oscillation parameters very precisely, and therefore, some of these models can be excluded in the future by these experiments. We have estimated the prediction of these models numerically and then used them as input to scrutinize these models in the neutrino experiments. Assuming the future best-fit values of $\theta_{23}$ and $\delta_{\rm CP}$ remain the same as the current one, our results show that at $5 \sigma$ C.L, Model-A can be excluded by T2HK whereas Model-B can be excluded by both T2HK and DUNE. Model-C cannot be excluded by T2HK and DUNE at $5 \sigma$ C.L. Further; our results show that JUNO alone can exclude Model-B at an extremely high confidence level if the future best-fit of $\theta_{12}$ remains at the current-one. We have also identified the region in the $\theta_{23}$ - $\delta_{\rm CP}$ parameter space, for which Model-A cannot be separated from Model-B in T2HK and DUNE. 
 
\end{abstract}

\maketitle
\flushbottom

\section{INTRODUCTION}

The Standard Model (SM) of elementary particles and their interactions suggests that quarks and leptons exist in three generations or families. However, the prevailing reason behind this and the  hierarchical nature of fermion masses still remains a mystifying puzzle in particle physics. Additionally, the differences in the mixing patterns between quarks and leptons and whether they have any underlying principles are not yet understood, and together these issues make up the ``flavor problem''.  One possible explanation for the mixing patterns in the lepton sector is non-Abelian discrete flavor symmetries, which have been extensively studied in the past few decades. However, the small mixing structure in the quark sector does not seem to support non-Abelian discrete symmetries, and attempts to explain both sectors with discrete flavor symmetries are rather challenging. In this article, we will focus only on the lepton sector and try to see if we can probe such models with the neutrino oscillation experiments rather than attempting to address the unified solution to the flavor problem. The discrete symmetry approach to lepton flavor has two main features: (i) it predicts certain values of the neutrino oscillation parameters, and (ii) it establishes algebraic relationships between some of the mixing parameters. These relationships are known as lepton (or neutrino) mixing sum rules and have been illustrated in various references \cite{Antusch:2007rk,Gehrlein:2016wlc}. However, in this article, we try to elucidate the theoretical frameworks based on $A_4$ modular discrete symmetry group utilizing certain seesaw mechanisms.
The most popular seesaw mechanisms used to generate the light neutrino masses as well as definite  flavor structure in the neutrino sector  are: type-I \cite{mohapatra1980neutrino,Brdar:2019iem,Branco:2020yvs,Bilenky:2010zza}, which incorporates three singlet right-handed heavy neutrinos,  type-II, with the inclusion of a scalar triplet \cite{Gu:2006wj,Luo:2007mq,Antusch:2004xy,Rodejohann:2004cg,Gu:2019ogb,McDonald:2007ka}, and type-III \cite{Liao:2009nq,Ma:1998dn,Foot:1988aq,Dorsner:2006fx,Franceschini:2008pz,He:2009tf},  where a fermion triplet is added to the SM particle content. In these approaches, the masses of the new heavy particles are rather heavy and are beyond the access of the present or future  experiments. Many other alternative approaches were proposed, e.g., linear seesaw \cite{Ma:2009du,Hirsch:2009mx,Gu:2010xc}, inverse seesaw \cite{Das:2012ze,Arganda:2014dta,Ma:2015raa,Dias:2012xp,CarcamoHernandez:2019eme,Dev:2012sg,Dias:2011sq,Bazzocchi:2010dt,Panda:2022kbn},  where the new physics scale responsible for neutrino mass generation can be brought down to  TeV scale, at the expense  of the inclusion  of new additional  fermion fields ($S_i$), which are SM singlets. Another interesting  idea that has received a lot of attention in recent times is the application of modular symmetry \cite{Leontaris:1997vw,Kobayashi:2018vbk,feruglio2019neutrino,deAdelhartToorop:2011re},  where the usage of excess flavon fields can naturally be avoided. In this case, the Yukawa couplings, which are holomorphic functions of modulus $\tau$, perform the role of flavons. Here $\tau$ is a complex variable that appears in the Dedekind eta function $\eta(\tau)$ \cite{Apostol1990}. When this modulus acquires the vacuum expectation value (VEV), it breaks the flavor symmetry. There exist plentiful works in the literature based on modular groups $S_3$ \cite{Mishra:2020gxg,Okada:2019xqk}, $S_4$ \cite{Penedo:2018nmg,Novichkov:2018ovf,Okada:2019lzv,Nomura:2023dgk}, $A_4$ \cite{Abbas:2020vuy,Nomura:2023kwz,Kim:2023jto,Kashav:2022kpk,Nagao:2020snm,Asaka:2020tmo,Nomura:2020opk,Okada:2020dmb,Behera:2020lpd,Ding:2019zxk,Altarelli:2005yx,Kashav:2021zir,Devi:2023vpe,Singh:2023jke,Kikuchi:2023jap,Petcov:2022fjf,Abbas:2022slb,Du:2022lij,Ding:2022bzs,Nomura:2021pld,Kuranaga:2021ujd},  $A_5$ \cite{Novichkov:2018nkm,Yao:2020zml}, double covering of $A_4$ \cite{Liu:2019khw,Mishra:2023cjc,Benes:2022bbg,Okada:2022kee}, $S_4$ \cite{Abe:2023qmr,Abe:2023ilq} and $A_5$ \cite{Wang:2020lxk,Behera:2022wco,Behera:2021eut}. These modular groups are quite successful in accommodating the observed  neutrino oscillation parameters. 

Due to the predictive features of modular symmetry group-based models, they can be probed in the forthcoming neutrino oscillation experiments.  In this paper, we consider three such models and study their viability in the upcoming neutrino experiments T2HK \cite{Hyper-Kamiokande:2016srs}, DUNE \cite{DUNE:2020ypp} and JUNO \cite{JUNO:2015zny}. The experiments T2HK and DUNE are the future accelerator-based long-baseline experiments, whereas JUNO is the medium baseline reactor experiment which is expected to start taking data in early 2024. The experiments T2HK and DUNE are expected to precisely measure the leptonic CP phase $\delta_{\rm CP}$ as well as the octant of the atmospheric mixing angle  $\theta_{23}$. The experiment JUNO will further improve the measurement of $\theta_{12}$. It is also expected that DUNE and JUNO experiments will determine the true nature of neutrino mass ordering. Also, T2HK will determine mass ordering at good significance by combining beam and atmospheric data. Currently, two orderings of the neutrino masses are allowed based on the sign of the atmospheric mass squared difference $\Delta m^2_{31}$. The positive sign of $\Delta m^2_{31}$ gives rise to normal mass ordering, whereas the negative sign of the $\Delta m^2_{31}$ gives rise to inverted mass ordering. The rest of the parameters i.e., $\theta_{13}$ and $\Delta m^2_{21}$ are currently very well measured \cite{Esteban:2020cvm}. Here it should be mentioned that the current measurements of $\theta_{23}$ and $\delta_{\rm CP}$ are very weak. Due to this, a large number of models are currently allowed, which predict a wide range of values regarding these two parameters. However, with the future measurements of these parameters by T2HK and DUNE, we expect to rule out many such models.  One way to test these models in the neutrino experiments is to use the sum rules which connects the model parameters to the leptonic mixing parameters \cite{Blennow:2020snb,Blennow:2020ncm,Chatterjee:2017ilf,Agarwalla:2017wct,Ballett:2016yod}. There are also studies which derived sum rules for the modular symmetry models \cite{Gehrlein:2020jnr} and tested them in the neutrino experiments \cite{Gehrlein:2022nss}. However, for the models that we consider in our study, it is not possible to derive the sum rules. In this case we numerically calculate the predicted values of the leptonic mixing parameters from these models and use them as an input while studying these models in the neutrino experiments. To the best of our knowledge this is the first work of this kind. In our work, we choose three different models in such a way that their prediction of $\theta_{23}$ and $\delta_{\rm CP}$ are very different from each other, but still, they are allowed because of the current wide allowed ranges of these two parameters. Therefore, the main motivation of our work is to see how well future neutrino experiments will be able  to constrain these models. Among these three models, we find that one of the models predicts a very narrow range of $\theta_{12}$. This provides us an opportunity to constrain this model using JUNO. In addition, we also notice that two of the models have partially common allowed parameter space in terms of values of $\theta_{23}$ and $\delta_{\rm CP}$. In our work, we study the capability of DUNE and T2HK to separate these two models.

The structure of this paper is as follows. In Sec. \ref{sec:models}, we provide an overview of the theoretical framework for the three models borrowed from the original work as cited in respective subsections \ref{sub:linear}, \ref{sub:type-1} and \ref{sub:type-III}. Further, in sec. \ref{sec:predicitions}, we demonstrate the allowed parameter space in terms of all six neutrino oscillation parameters for these three models illustrated in \ref{sec:NA_model-A}, \ref{sec:NA_model-B} and \ref{sec:NA_model-C} respectively. In Sec.~\ref{sec:exp_char}, we highlight the features and characteristics of T2HK and DUNE, which we use in our analysis. In Sec.~\ref{sec:results_simulation}, we present our results, and finally, in Sec.~\ref{sec:conclusion}, we pen down our ascertainment in conclusion.
\section{Model Framework}
\label{sec:models}

In this section, we present an overview of the models employed in our analysis, which are based on modular symmetry. Specifically, we focus on the utilization of $A_4$ modular symmetry to preserve the invariance of the superpotential. Our models incorporate different seesaw mechanisms to account for the tiny masses of neutrinos. Here, we provide a description of these seesaw models, including the particle content, charges, and modular weights under $A_4$ modular symmetry.
\subsection{Linear Seesaw mechanism with $A_4$ Modular symmetry}
\label{sub:linear}

The model considered here, which we referred to as Model-A, involves the extension of the Standard Model with an additional  discrete $A_4$ modular symmetry. Broadly, it can be considered a modified version of  a type-I seesaw that follows the flavor structure of a linear seesaw. To enrich the particle spectrum, six extra singlet heavy  superfields ($N_{R_i}$ \& $S_{L_i}$) along with a weighton field ($\rho_a$) are introduced. The extra super-multiplets $N_{R_i}$ and $S_{L_i}$ transform as  triplets under the $A_4$ modular group, while $\rho_a$ transforms as a singlet. However, the modular Yukawa couplings are also charged under $A_4$ symmetry being a triplet with modular weight $k_I=2$. The modular symmetry restricts the use of excessive flavon fields, which would otherwise overpopulate the particle spectrum and reduce the model's predictability when working in BSM. This is achievable because the Yukawa couplings acquire a modular form and take on the role of extra flavon fields and is applicable to subsequent models discussed in this work. The particle spectrum clubbed with modular Yukawa couplings of the model and their group charges under various  symmetries, along with their modular weights,  are depicted in Table \ref{tab:fields-linear}. Generally,  the linear seesaw framework can be realized by assigning suitable  charges under different groups to the comprising particles. 

Additionally, the model has an additional $U(1)_X$ global symmetry as outlined in the original paper \cite{Behera:2020sfe}, enforced to eliminate certain undesirable terms in the superpotential.  The $A_4$  symmetry is believed to be broken at a scale much higher than the electroweak symmetry breaking \cite{Dawson:2017ksx}. The masses of the additional superfields are generated by assigning a non-zero vacuum expectation value to the singlet weighton.  

\begin{center} 
\begin{table}[htpb]
\centering
\begin{tabular}{||c||c|c|c|c|c|c||c|c||c||}\hline\hline  
  & \multicolumn{6}{c||}{Fermions} & \multicolumn{2}{c||}{Scalars} & \multicolumn{1}{c||}{Yukawa couplings} \\ \hline \hline

Fields & ~$e_R^c$~& ~$\mu_R^c$~  & ~$\tau_R^c$~& ~$L_{\wp L}$~& ~$N_R$~& ~$S_L^c$~& ~$H_{u,d}$~&~$\rho_a$ & $\textbf{
Y}=(y_1, y_2, y_3)$\\ \hline 
$SU(2)_L$ & $1$ & $1$ & $1$ & $2$ & $1$ & $1$ &$2$&$1$&$-$ \\ \hline
$U(1)_Y$ & $1$ & $1$ & $1$ & $-\frac{1}{2}$ & $0$ &$0$& $\frac{1}{2}, -\frac{1}{2}$ &$0$&$-$  \\ \hline
$U(1)_X$ & $1$ & $1$ & $1$ & $-1$ & $1$ & $-2$ & 0 &$1$ & $-$  \\ \hline
$A_4$ & $1$ & $1'$ & $1''$ & $1, 1^{\prime \prime}, 1^{\prime }$ & $3$ & $3$ & $1$ & $1$  & $3$ \\ \hline

$k_I$ & $1$ & $1$ & $1$ & $-1$ & $-1$ & $-1$ & $0$  & $0$ & $2$\\ 
\hline \hline
\end{tabular}
\caption{Particle content and modular Yukawa couplings of the model and their charges under $SU(2)_L \times U(1)_Y \times U(1)_{X} \times A_4 $ where $k_I$ is the number of modular weight.}
\label{tab:fields-linear}
\end{table}
\end{center}
The invariant super-potential in the present model framework   is represented 
as:
\begin{eqnarray}
\mathcal{W} &=&  y_{\ell_{}}^{\wp \wp}  {L}_{\wp_L} H_d ~\wp_R^c + G_D   {L}_{\wp_L} H_u~ (\bm{Y} N_R)_{1, 1^{\prime \prime}, 1^\prime} +  G^\prime_D \left[   {L}_{\wp_L} H_u~ (\bm{Y} S^c_L)_{1, 1^\prime, 1^{\prime \prime}} \right]\frac{\rho_a^3}{\Lambda_a^3} \nonumber \\ 
&+& [\alpha_{NS} \bm{Y} ({S^c_L} N_R)_{\rm sym} + \beta_{NS} \bm{Y} ({S^c_L} N_R)_{\rm Anti-sym} ]\rho_a\;,
\label{super_p}
\end{eqnarray}
where, $G_D = {\rm diag} (\alpha_D,~\beta_D,~\gamma_D)$ and $G_D^\prime = {\rm diag} (\alpha_D^\prime,~\beta_D^\prime,~\gamma_D^\prime)$ being the diagonal matrices containing six free parameters,  $\alpha_{NS}$ and $\beta_{NS}$ are other free parameters involved in the super-potential and $\wp = (e,~\mu,~\tau)$ corresponding to the lepton flavors and $\Lambda_a$ is the cut-off scale.

Depending upon the charge assignment to left-handed (LH) and right-handed (RH) charged leptons, which are singlets under $A_4$,   the charged lepton mass matrix obtained from the first term in Eq.(\ref{super_p}), is found to be diagonal, and the couplings can be adjusted to achieve the observed charged lepton masses. The mass matrix takes the form
\begin{align}
M_\ell = \begin{pmatrix}  y_{\ell_{}}^{ee} v_d/\sqrt{2}  &  0 &  0 \\
 0  &  y_{\ell_{}}^{\mu \mu} v_d/\sqrt{2}  &  0 \\
0  &  0  &  y_{\ell_{}}^{\tau \tau} v_d/\sqrt{2}        \end{pmatrix}  = \begin{pmatrix}  m_e  &  0 &  0 \\
0  &  m_\mu  &  0 \\
0  &  0  &  m_\tau      \end{pmatrix}.                 
\label{Eq:Mell} 
\end{align}
Here, $m_e$, $m_\mu$, and $m_\tau$ are the observed charged lepton masses. Further, in Eq.(\ref{super_p}) the second term involves $N_{R_i}$, which being an $A_4$ triplet,  its contraction happens with triplet modular Yukawa coupling \textbf{$Y$}, hence resulting  the Dirac neutrino mass matrix shown below
\begin{align}
M_D&=\frac{v_u}{\sqrt2}
\left[\begin{array}{ccc}
\alpha_D & 0 & 0 \\ 
0 & \beta_D & 0 \\ 
0 & 0 & \gamma_D \\ 
\end{array}\right]
\left[\begin{array}{ccc}
y_1 &y_3 &y_2 \\ 
y_2 &y_1 &y_3 \\ 
y_3 &y_2 &y_1 \\ 
\end{array}\right]_{LR}.                   
\label{Eq:Mell} 
\end{align}
Similarly, the flavor structure for the pseudo-Dirac term is due to the contraction of $A_4$ triplet $S_{L_i}$ with modular Yukawa coupling \textbf{$Y$}, and its mass matrix takes the form,
\begin{align}
M_{LS}&=\frac{v_u}{\sqrt2}\left(\frac{v_{\rho_a}}{\sqrt{2}\Lambda_a}\right)^3
\left[\begin{array}{ccc}
\alpha^\prime_D & 0 & 0 \\ 
0 & \beta^\prime_D & 0 \\ 
0 & 0 & \gamma^\prime_D \\ 
\end{array}\right]
\left[\begin{array}{ccc}
y_1 &y_3 &y_2 \\ 
y_2 &y_1 &y_3 \\ 
y_3 &y_2 &y_1 \\ 
\end{array}\right]_{LR},                  
\label{Eq:Mell} 
\end{align}\\
where, $v_u$, $v_d$ and $v_{\rho_a}$ are the VEV of $H_u$, $H_d$ and $\rho_a$ respectively. Finally, the mixing between the heavy fermions $N_{R_i}$ and $S_{L_i}$ results in a mass matrix consisting of both symmetric and anti-symmetric matrices with $\alpha_{NS} \gg \beta_{NS}$ as given below,
\begin{align}
M_{RS}&=\frac{v_{\rho_a}}{\sqrt2}
 \left(
 \frac{\alpha_{NS}}{3}\left[\begin{array}{ccc}
2y_1 & -y_3 & -y_2 \\ 
-y_3 & 2y_2 & -y_1 \\ 
-y_2 & -y_1 & 2y_3 \\ 
\end{array}\right]_{sym}
+
\beta_{NS}
\left[\begin{array}{ccc}
0 &y_3 & -y_2 \\ 
-y_3 & 0 & y_1 \\ 
y_2 & -y_1 &0 \\ 
\end{array}\right]_{asym}
\right). \label{yuk:MRS}
\end{align}

The linear seesaw is a modified form of type-I seesaw, and the mass formula for  light neutrinos in this framework  is given by
\begin{equation}
    m_\nu = -M_D M_{RS}^{-1} M_{LS}^T ~+~ {\rm transpose}\;,
    \label{lin-mass}
\end{equation}
with $M_{RS} \gg M_D> M_{LS}$. It should be emphasized here that, we have neglected the term proportional to $\left (\frac{S_L^cS_L^c\rho^4}{\Lambda^3}\right )$ in eqn.(\ref{super_p}) by assuming the corresponding multiplicative  free parameter to be extremely small, in order to retain the linear seesaw neutrino mass structure. 
\subsection{Type-I seesaw without any flavon fields}
\label{sub:type-1}
Here, we consider another model based on the $A_4$ modular symmetry proposed by Feruglio  \cite{feruglio2019neutrino} identified as Model-B. In this model, the Higgs\footnote{The super-multiplet's (i.e., $H_u$ \& $H_d$) VEV are related to SM Higgs VEV by the relation $v_H = \frac12 \sqrt{v_u^2 + v_d^2}$, and $\tan \beta = (\nicefrac{v_u}{v_d}).$} and lepton super-fields undergo specific transformations, which are detailed in Table \ref{tab:Model-2}. This model is free from extra flavon fields and only includes three right-handed (RH) neutrinos.\\
\begin{table}[h!]
\begin{center}
\begin{tabular}{||c||c|c|c|c|c||c||c||} \hline\hline 
& \multicolumn{5}{c||}{Fermions} & \multicolumn{1}{c||}{Scalars} & \multicolumn{1}{c||}{Yukawa couplings} \\ \hline \hline
Fields & ~$E^c_{1R}$~& ~$E^c_{2R}$~  & ~$E^c_{3R}$~& ~$L$~& ~$N^c_{R}$~&~ $H_{u,d}$ & ~$\textbf{
Y}=(y_1, y_2, y_3)$~\\ \hline \hline
$SU(2)_L$ & $1$ & $1$ & $1$ & $2$ & $1$ & $2$ & $-$ \\ \hline
$U(1)_Y$ & $1$ & $1$ & $1$ & $-\frac{1}{2}$ & $0$ & $\frac{1}{2}, -\frac{1}{2}$ &$ -$ \\ \hline
$A_4$ & $1$ & $1''$ & $1'$ & $3$ & $3$ & $1$ & $3$ \\ \hline
$k_I$ & $-1$ & $-1$ & $-1$ & $-1$ & $-1$ & $0$ & $2$  \\ \hline

\hline \hline
\end{tabular}
\caption{Particle content and Yukawa couplings under $SU(2)_L \times U(1)_Y \times A_4$ modular symmetry, where, $k_I$ being the modular weight.}
\label{tab:Model-2}
\end{center}
\end{table}

The super-potential of the model can be expressed as
\begin{eqnarray}
    \mathcal{W} = w_e + w_\nu \;,
    \end{eqnarray} \vspace{-1cm}
with
\begin{eqnarray}
   && w_e = \alpha_l E_1^c H_d (LY)_{1} + \beta_l E_2^c H_d (LY)_{1^\prime} +\gamma_l E_3^c H_d (LY)_{ 1^{\prime \prime}} \;,\\
   \hspace{-0.5cm}
   {\rm and}~~~~~~~~~~~~~~~
    &&w_\nu = g(N^c_R H_u L Y)_1 + M_b (N^c_R N^c_R Y)_1\;,\label{NRc}
\end{eqnarray}
where, $g, \alpha_l, \beta_l, \gamma_l$ are the free parameters and  $M_b$ is the  mass parameter. The mass matrix for the charged lepton sector is a non-diagonal matrix because the left and right-handed charged leptons are triplets and singlets, and, due to the product rule of $A_4$ symmetry, they attain a form given below.
\begin{eqnarray}
    M_L &=& \frac{v_d}{\sqrt{2}}\begin{pmatrix}
        \alpha_l y_1 & \alpha_l y_3 & \alpha_l y_2 \\
        \beta_l y_2 & \beta_l y_1 & \beta_l y_3 \\
        \gamma_l y_3 & \gamma_l y_2 & \gamma_l y_1
    \end{pmatrix},
\end{eqnarray}
where $v_d$ is the VEV of $H_d$, and the free parameters can be determined by using the below identities:
 \begin{eqnarray}
   \mathrm{ Det}(M_L M_L^{\dagger}) &=& m^2_e m^2_\mu m^2_\tau, \nonumber \\
    \mathrm{ Tr}(M_L M_L^{\dagger}) &=& m^2_e+ m^2_\mu+ m^2_\tau, \nonumber \\
    \frac{1}{2}\left ( [\mathrm{Tr}(M_L M_L^{\dagger})]^2 -[\mathrm{Tr}(M_L M_L^{\dagger})^2] \right ) &=&  m^2_e m^2_\mu+ m^2_\mu m^2_\tau + m^2_\tau m^2_e\;,
 \end{eqnarray}
 where, $m_e$, $m_\mu$ and $m_\tau$ are the  charged lepton masses \cite{Zyla:2020zbs}. Further, the Dirac mass matrix emerges as given below due to contraction between LH lepton doublets, RH neutrinos, and modular Yukawa couplings as they are $A_4$ triplets.
 \begin{eqnarray}
    M_D = \frac{v_u}{\sqrt{2}} \begin{pmatrix}
         2\mathsf{g}_1 y_1 & (-\mathsf{g}_1 + \mathsf{g}_2) y_3 & (-\mathsf{g}_1 -\mathsf{g}_2) y_2 \\
         (-\mathsf{g}_1-\mathsf{g}_2) y_3 & 2\mathsf{g}_1y_2 & (-\mathsf{g}_1+\mathsf{g}_2) y_1\\
         (-\mathsf{g}_1+\mathsf{g}_2) y_2
         & (-\mathsf{g}_1 -\mathsf{g}_2) y_1 & 2\mathsf{g}_1 y_3
     \end{pmatrix}.
 \end{eqnarray}

where, $v_u$ is the VEV of $H_u$. Finally, the Majorana mass matrix for  heavy right-handed neutrinos obtained from the second term of eqn.(\ref{NRc}), expressed as
\begin{eqnarray}
  M_R =  \begin{pmatrix}
        2y_1 & -y_3 & -y_2 \\
        -y_3 & 2y_2 & -y_1 \\
        -y_2 & -y_1 & 2y_3
    \end{pmatrix} M_b\;.
\end{eqnarray}
As it is a type-I seesaw mechanism hence, the light neutrino mass formula is given by 
\begin{equation}
    m_\nu = - M_D M_R^{-1} M_D^T\;.
\end{equation}

 \subsection{Type-III seesaw }
\label{sub:type-III}
 \begin{table}[htbp]
\begin{center}
\begin{tabular}{||c||c|c|c|c|c||c|c||c||} \hline\hline 
& \multicolumn{5}{c||}{Fermions} & \multicolumn{2}{c||}{Scalars} & \multicolumn{1}{c||}{Yukawa couplings} \\ \hline \hline
Fields & ~$E^c_{1R}$~& ~$E^c_{2R}$~  & ~$E^c_{3R}$~& ~$L$~& ~$\Sigma^c_{R_i}$~&~ $H_{u,d}$~&~$\rho_c$~&~ $\textbf{Y}=(y_1, y_2, y_3)$ \\ \hline \hline
$SU(2)_L$ &$1$&$1$&$1$&$2$&$3$&$2$&$1$&$-$ \\ \hline
$U(1)_Y$ &$1$&$1$&$1$&$-\frac{1}{2}$&$0$&$\frac{1}{2},-\frac{1}{2}$&$0$&$-$ \\ \hline
$U(1)_{B-L}$ &$1$&$1$&$1$&$-1$&$1$&$0$&$-2$&$-$ \\ \hline
$A_4$ & $1$ & $1'$ & $1''$ & $1, 1^{\prime \prime}, 1^{\prime }$ & $3$ & $1$ & $1$& $3$ \\ \hline
$k_I$ & $0$ & $0$ & $0$ & $0$ & $-2$ & $0$ & $2$ & $2$ \\ \hline

\hline
\end{tabular}
\caption{Particle content and Yukawa couplings of the type-III seesaw and their charges under $SU(2)_L \times U(1)_Y \times U(1)_{B-L} \times A_4 $, and $k_I$ is the modular weight.}
\label{tab:Model3}
\end{center}
\end{table}

The  model considered here, referred to as Model-C, employs a type-III seesaw mechanism with $A_4$ modular symmetry and $U(1)_{B-L}$ symmetry in addition to SM symmetries. The model comprises BSM particles, i.e., three right-handed fermion triplets $(\Sigma_{R_i})$ and weighton $(\rho_c)$. Hence, $A_4$ symmetry and $U(1)_{B-L}$ symmetries are broken at a significantly high scale than the electroweak symmetry-breaking scale. Table (\ref{tab:Model3}) provides the $A_4$ charges and modular weights for the added particles with elaborated discussion present in original work \cite{Mishra:2022egy}. The purpose of incorporating extra $U(1)_{B-L}$ symmetry is to avoid undesirable terms in the superpotential, which cannot be prevented by $A_4$ modular symmetry alone. The non-zero VEV of the singlet weighton aids in providing mass to the heavy RH neutrinos. The complete superpotential of the model is given as
\begin{eqnarray}
 \mathcal{W}
                   &=& - y_{ij}E^c_{R_i}H_d L_j   - ~ (\alpha_\Sigma)_{ij} \left[{H_u} \Sigma^c_{R_i} \sqrt{2}\textbf{Y} L_j \right] \nonumber\\ && -\frac{M^{\prime}_{\Sigma_{}}}{2} \left(\beta_{\Sigma} {\rm{Tr}}  \left[\mathbf{{\Sigma^c_{R_i}}}\textbf{Y} \mathbf{{\Sigma^c_{R_i}}}\right]_{\rm{s}} + \gamma^\prime_\Sigma {\rm{Tr}} \left[\mathbf{{\Sigma^c_{R_i}}}\textbf {Y} \mathbf{ {\Sigma^c_{R_i}}}\right]_{\rm a}\right) \frac{\rho_c}{\Lambda_c}  \;, 
                    \label{charged lepton}
\end{eqnarray}
where, $\alpha_{\Sigma}$ = diag $(a_1,~a_2,~a_3)$, $\beta_{\Sigma}$ = diag $(b_1,~b_2,~b_3)$  and $\gamma^\prime_{\Sigma} = diag (\gamma_1, \gamma_2, \gamma_3)$ representing free parameter diagonal matrices, with $M^{\prime}_{\Sigma_{}}$ as the free mass parameter and $\Lambda_c$ is the cut-off scale with the charged lepton mass matrix being diagonal, similar to, as depicted in Model-A. The interaction between the neutral multiplet of a fermion triplet and the LH neutrino leads to the Dirac mass term, which determines the Dirac mass matrix:

\begin{align}
    M_D&=v_u
\left[\begin{array}{ccc}
a_1 & 0 & 0 \\ 
0 & a_2 & 0 \\ 
0 & 0 & a_3 \\ 
\end{array}\right]
\left[\begin{array}{ccc}
y_1 &y_3 &y_2 \\ 
y_2 &y_1 &y_3 \\ 
y_3 &y_2 &y_1 \\ 
\end{array}\right],    \label{eqn:Dirac-M3}
\end{align}\
where, $v_u$ is the VEV of $H_u$ and the Majorana mass matrix is given as,
\begin{align}
M_{R}&=\frac{v_{\rho_c}}{\Lambda_c\sqrt2} \left(\frac{M^{\prime}_{\Sigma}}{2}\right) \left( \frac{\beta_{\Sigma}}{3}\left[\begin{array}{ccc}
2y_1 & -y_3 & -y_2 \\ 
-y_3 & 2y_2 & -y_1 \\ 
-y_2 & -y_1 & 2y_3 \\ 
\end{array}\right]_{sym} + \gamma_{\Sigma}^\prime \left[\begin{array}{ccc}
0 & y_3 & -y_2 \\ 
-y_3 & 0 & y_1 \\ 
y_2 & -y_1 & 0 \\ 
\end{array}\right]_{asym} \right),
\label{Majorana-M3}
\end{align}\
where, $v_{\rho_c}$ is the VEV of $\rho_c $. The mass matrix for active neutrinos within the type-III seesaw framework is expressed as follows:
\begin{align}
m_\nu&= - M_D M_R^{-1} M_D^T\;.
\label{nmass}
\end{align}
It is worth to emphasize that
in this work, we mainly focus on three distinct flavor models based on $A_4$ modular symmetry, wherein these models incorporate different seesaw mechanisms to account for the light neutrino masses. Model-A is based on linear seesaw mechanism, where the light neutrino mass matrix is given in eqn. (\ref{lin-mass}). It requires interaction terms of dimensions 4 and 7 to implement linear seesaw formalism in this framework  as seen from eqn. (\ref{super_p}). On the other hand, Model-B incorporates type-I seesaw mechanism, so the interaction terms considered up to dim-4 are sufficient for studying neutrino phenomenology. The contribution arising from the Weinberg operator, which violates the Lepton number by two units, is neglected in our analysis. In model-C, type-III seesaw mechanism is considered where the SM gauge symmetry is extended by including $U(1)_{B-L}$ symmetry in addition to the $A_4$ modular symmetry.  The terms up to dimension-5, 
are utilized to generate Majorana masses for neutrinos, preserving their conservation under the $U(1)_{B-L}$ symmetry.
Since different seesaw mechanisms are utilized in these models to generate the light neutrino masses, it is impractical to treat them on equal footing by keeping  the interaction terms of same dimensions in all the three models.

\begin{table}[htbp]
    \centering
    \begin{tabular}{|c|c|c|c|c|c|c|}
    \hline \hline
       Osc. Parameters   &   $\sin^2\theta_{12}$ ~&~$\sin^2\theta_{23}$~&~$\sin^2\theta_{13}$~&~$\Delta m_{21}^2$ [eV$^2$]~&~$| \Delta m_{31}^2$ |[eV$^2$]~&~$\delta_{\rm CP}$\\ \hline \hline
         Nu-Fit (NH) & 0.303&0.451&0.02225&7.41 $\times 10^{-5}$& 2.507 $\times 10^{-3}$&232$^\circ$ \\ \hline
          Nu-Fit (IH) & 0.303&0.569&0.02223&7.41 $\times 10^{-5}$& 2.486 $\times 10^{-3}$&276$^\circ$ \\ \hline
         
          Model-A & 0.315&0.428&0.02343&7.65 $\times 10^{-5}$& 2.502 $\times 10^{-3}$& 45.30$^\circ$ \\ \hline
          Model-B & 0.340 &0.452&0.02138&7.73 $\times 10^{-5}$& 2.474 $\times 10^{-3}$&0$^\circ$ \\ \hline
          Model-C & 0.308 &0.469&0.02169&7.17 $\times 10^{-5}$& 2.461 $\times 10^{-3}$&213.18$^\circ$ \\ \hline
          
    \end{tabular}
    \caption{In the above table, we elucidate the best-fit value for the oscillation parameters (i.e., mixing angles $\sin^2 \theta_{12}, \sin^2 \theta_{23}$ $\&$ $\sin^2 \theta_{13} $ and mass squared differences ($\Delta m_{21}^2$ and $\Delta m_{31}^2$) and $\delta_{\rm{CP}}$ phase.) }
    \label{tab:Nufit}
\end{table}

 \section{Prediction of the leptonic mixing parameters from the models} 
 \label{sec:predicitions}
 
 This section is accentuated for discussing the results obtained for three different models whose theoretical frameworks are aforementioned in Sec. \ref{sec:models}. In order to conduct numerical analysis, we are utilizing the NuFIT neutrino oscillation data from \cite{website,Esteban:2020cvm} at a 3$\sigma$ interval as follows:
\begin{align}
{\rm NO}:& \sin^2\theta_{13}=[0.02052, 0.02398],\ 
\sin^2\theta_{23}=[0.408, 0.603],\ 
\sin^2\theta_{12}=[0.270, 0.341],\nn \\
& \Delta m^2_{\rm 31}=[2.427, 2.59]\times 10^{-3}\ {\rm eV}^2,\
\Delta m^2_{\rm 21}=[6.82, 8.03]\times 10^{-5}\ {\rm eV}^2,
\label{eq:mix-NO}\\\nn \\
{\rm IO}: &\sin^2\theta_{13}=[0.02048, 0.02416],\
\sin^2\theta_{23}=[0.412, 0.613],\ 
\sin^2\theta_{12}=[0.270, 0.341],\nn \\
&| \Delta m^2_{\rm 31} | =[ 2.406,2.57]\times 10^{-3}\ {\rm eV}^2,\
\Delta m^2_{\rm 21}=[6.82, 8.03]\times 10^{-5}\ {\rm eV}^2.\label{eq:mix-IO}
\end{align}
Here, we numerically diagonalize the neutrino mass matrices for each model through the relation $U^\dagger {\cal M}U= {\rm diag}(m_1^2,~m_2^2,~m_3^2)$, where  ${\cal M}=m_\nu m_\nu^\dagger$ and $U$ is a unitary matrix\footnote{For non-diagonal charged lepton sector the lepton mixing matrix is given as $U'= U_l^\dagger U$, where, $U_l$ matrix diagonalizes the charged lepton mass matrix \cite{Hochmuth:2007wq}.}  from which the neutrino mixing angles can be extracted  using the standard relations:
\begin{eqnarray}
\sin^2 \theta_{13}= |U_{13}|^2,~~~~~\sin^2 \theta_{23}= \frac{|U_{23}|^2}{1-|U_{13}|^2},~~~~\sin^2 \theta_{12}= \frac{|U_{12}|^2}{1-|U_{13}|^2}\;.
\end{eqnarray}

The input parameters are randomly scanned over for each model individually, and presented in  Tables \ref{tab:M1-model-para}, \ref{tab:M2-model-para} and \ref{tab:M3-model-para} for Model-A, Model-B, and Model-C, respectively. The parameter space for the allowed regions is initially filtered by the observed $3\sigma$ limit of solar and atmospheric mass squared differences and further constrained by mixing angles 
and the observed sum of active neutrino masses $0.058 \leq\Sigma m_{\nu_i} \leq 0.12 ~\rm eV$ \cite{Planck:2018vyg, Vagnozzi:2017ovm} ($i$=1,2,3),  for Model-A and Model-C as these models satisfy normal hierarchy and  $0.098 \leq \Sigma m_{\nu_i} \leq 0.12 ~\rm eV$ \cite{RoyChoudhury:2018gay} for Model-B, which follows the inverted hierarchy. The best-fit values of the input parameters are obtained by utilizing the chi-square minimization technique and the general chi-square formula \cite{Roe:2015fca, Ding:2021eva},
\begin{align}
    \chi ^2 = \sum_i \left ( \frac{T_i(z)-E_i}{\sigma_i}      \right )^2 ,
\end{align}
where $E_i$ is the experimentally observed  best-fit value of oscillation parameters from NuFIT and $T_i(z)$ is the theoretical prediction  for the corresponding oscillation parameter as a function of $z$, where $z$ indicates input parameters in the model, while $\sigma_i$ is the $1\sigma$ errors in $E_i$.  

By examining the values of the input parameters within their respective ranges, we calculate the $\chi^2$ for all available observables in the neutrino sector, which include two mass-squared differences, three mixing angles, and CP phase ($\delta _{\rm CP}$). These calculations yield a cumulative $\chi^2$ minimum, which we use to determine the values of the free parameters that correspond to the minimum, also known as the best-fit values. By employing these best-fit values of the input parameters, we can refine the ranges for each model parameter by aligning them closely around these values. By adopting this approach, we can generate model predictions by placing emphasis on the parameter value that is most likely. Next, we highlight the nitty-gritty of the model parameter space for each case below.

\subsection{Model-A}
\label{sec:NA_model-A}

To fit the current neutrino oscillation data, we chose the  ranges for  the model parameters as shown in Table \ref{tab:M1-model-para},

\begin{table}[htpb]
    \centering
\begin{tabular}{|c|c||c|c|} \hline \hline 
 Parameter & Range
& Parameter & range  \\ \hline\hline 
${\rm Re}[\tau]$ & $[-0.5,0.5] $ &  $\beta^\prime_{D} $&[$6.0-9.9$]$\times 10^{-3}$ 
\\ \hline 
${\rm Im}[\tau]$&$[1.2,1.8]$ &  $\gamma^\prime_{D}$& [$3.2-8.3$]$\times 10^{-3}$\\ \hline $\alpha_{D} $  &  [$6.7-9.7$]$\times 10^{-6}$ & $\alpha_{NS}$ & [$0.1-0.24$] \\ \hline $\beta_{D}$  
    & [$4.4-4.8$]$\times 10^{-6}$ & $\beta_{NS}$& [$1-2.2$]$\times 10^{-5}$\\ \hline $\gamma_{D}$ 
  & [$5.2-8.8$]$\times 10^{-6}$ & $\Lambda_a $&$ [10^5, 10^6]~ \rm GeV $ \\ \hline $\alpha^\prime_{D} $ & [$2.3-6.3$]$\times 10^{-3}$ & $v_{\rho_a} $&$ [10^4, 10^5]~ \rm GeV $\\ \hline

\end{tabular}
\caption{Model parameters ranges for Model-A  }
    \label{tab:M1-model-para}
\end{table}

 The prediction of the leptonic mixing parameters for Model-A is shown in the top row of Fig.~\ref{fig1}. From the panels, we see that the model allows all the values of the parameters $\theta_{12}$, $\theta_{13}$, $\theta_{23}$, $\Delta m^2_{21}$ and $\Delta m^2_{31}$ within the current $3 \sigma$ values of their global fit except $\delta_{\rm CP}$.  The allowed values of $\delta_{\rm CP}$ lies within the range of $0^\circ \le \delta_{\rm CP} < 89^\circ$. As the T2HK and DUNE will be able to measure $\delta_{\rm CP}$ very precisely, this model can be constrained in the future based on the data from these two experiments. The best-fit values of the oscillation parameters as obtained from this model are listed in the third row of Table~\ref{tab:Nufit}.

 \subsection{Model-B}
 \label{sec:NA_model-B}
 Similarly, we showcase the model parameter space in Table \ref{tab:M2-model-para} for Model B, which satisfies the neutrino experimental data. 
\begin{table}[htpb]
    \centering
\begin{tabular}{|c|c|c|c|c|c|} \hline
  $~~\rm Re(\tau)~~$  &  ~~$\rm Im(\tau)$~~  
    & ~~$g_1$ ~~
  &  ~~$g_2$ ~~& $~~M_b$ [GeV]~~  \\ \hline
[$-0.1, 0.1$] & [$1.5, 2.0$] & [$1, 3$]$\times 10^{-5}$ & [$2, 5$]$\times 10^{-5}$ & [$10^{6}, 10^{7}$] 
 \\ \hline
\end{tabular}
\caption{Ranges required for input parameters in Model-B}
    \label{tab:M2-model-para}
\end{table}
  The prediction of this model in terms of the neutrino oscillation parameters is shown in the middle row of Fig.~\ref{fig1}. For this model, the parameters $\theta_{13}$ and $\Delta m^2_{21}$ are unconstrained. For $\Delta m^2_{31}$ this model disallows a small region around the lower edge of the $3 \sigma$ bound for this parameter, i.e., $|\Delta m^2_{31}| < 2.43 \times 10^{-3} $. This model allows only $\delta_{\rm CP} = 0^\circ$. Additionally, this model predicts lower octant of $\theta_{23}$, i.e., $\sin^2\theta_{23} < 0.459$ and also a narrow range of $\theta_{12}$ with $\sin^2\theta_{12}$ around 0.34 which lies in the upper edge of the current $3\sigma$ according to NuFIT. Therefore, this model can be constrained by all three experiments that we consider in this present study. The prediction of $\theta_{23}$ and $\delta_{\rm CP}$ can be constrained by DUNE and T2HK, whereas the prediction of $\theta_{12}$ can be constrained by JUNO. Note that the allowed region of Model-B is basically a part of the allowed parameter space of Model-A in terms of parameter $\theta_{23}$ and $\delta_{\rm CP}$. Therefore it will be interesting to see the capability of DUNE and T2HK to distinguish these two models. The best-fit values of the oscillation parameters as obtained from this model are listed in the fourth row of Table~\ref{tab:Nufit}. 

Here it should be mentioned that Model-A is allowed for normal ordering of the neutrino masses, whereas Model-B is allowed for the inverted ordering. Therefore, once the true mass ordering of the neutrinos is determined, one of these models will be excluded immediately. However, in this study, we assume that both the mass ordering of the neutrinos are allowed, and we will study the distinguishability of these two models in DUNE and T2HK based on their prediction of the leptonic mixing parameters.
\subsection{Model-C}
\label{sec:NA_model-C}

Finally, the parameter space for Model-C, satisfying the ranges of the neutrino experimental bounds, is depicted in Table \ref{tab:M3-model-para}.
\begin{table}[htpb]
    \centering
\begin{tabular}{|c|c||c|c|} \hline \hline 
 Parameter & Range
& Parameter & range  \\ \hline\hline 
${\rm Re}[\tau]$ & $[-0.5,0.5] $ &  $b_{2} $&[$0.8-8$]$\times 10^{-1}$ 
\\ \hline 
${\rm Im}[\tau]$&[$0.75,2$] &  $b_{3}$& [$1-7$]$\times 10^{-3}$\\ \hline $a_1 $  &  [$5-10$]$\times 10^{-7}$ & $\gamma_{\Sigma}$ & [$0.1, 1$]$\times 10^{-9}$\\ \hline $a_{2}$  
    & [$4.5-10$]$\times 10^{-6}$ & $M^\prime_{\Sigma}$& [$10^{7}, 10^{8}$]~ \rm GeV \\ \hline $a_{3}$ 
  & [$0.5-5$]$\times 10^{-7}$ & $\Lambda_c $&$ [10^7, 10^8]~ \rm GeV $ \\ \hline $b_{1} $ & [$0.7-5$]$\times 10^{-2}$ & $v_{\rho_c} $&$ [10^6, 10^7]~ \rm GeV $\\ \hline

\end{tabular}
\caption{Ranges required for free parameters in Model-C}
    \label{tab:M3-model-para}
\end{table}

 \begin{figure}[htpb]
    \centering
    \subfloat[]{\includegraphics[height=45mm, width=54mm]{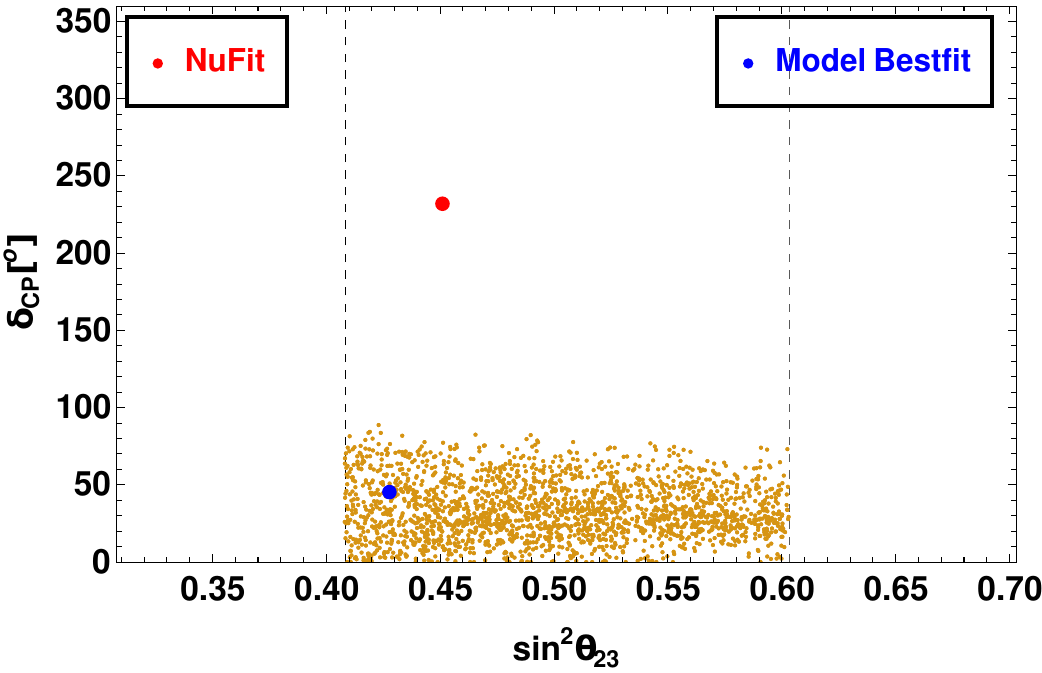}\label{ss23_dcp_model_A}}
    \subfloat[]{\includegraphics[height=45mm, width=54mm]{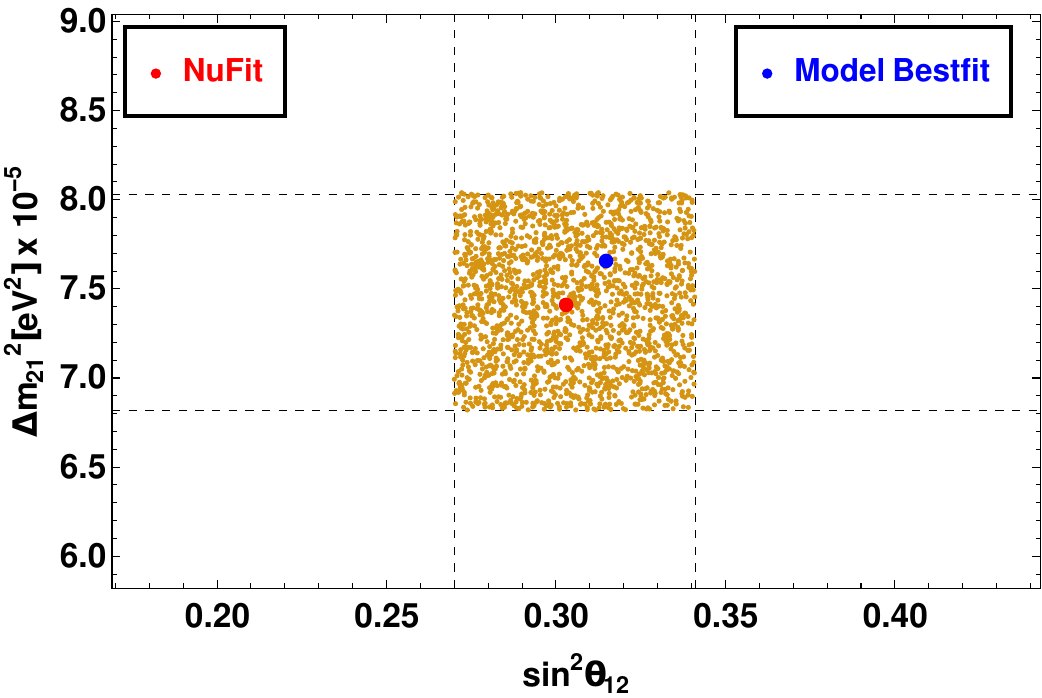}\label{ss12_dm12_model_A}}
    \hspace{0.1 cm}
    \subfloat[]{\includegraphics[height=45mm, width=54mm]{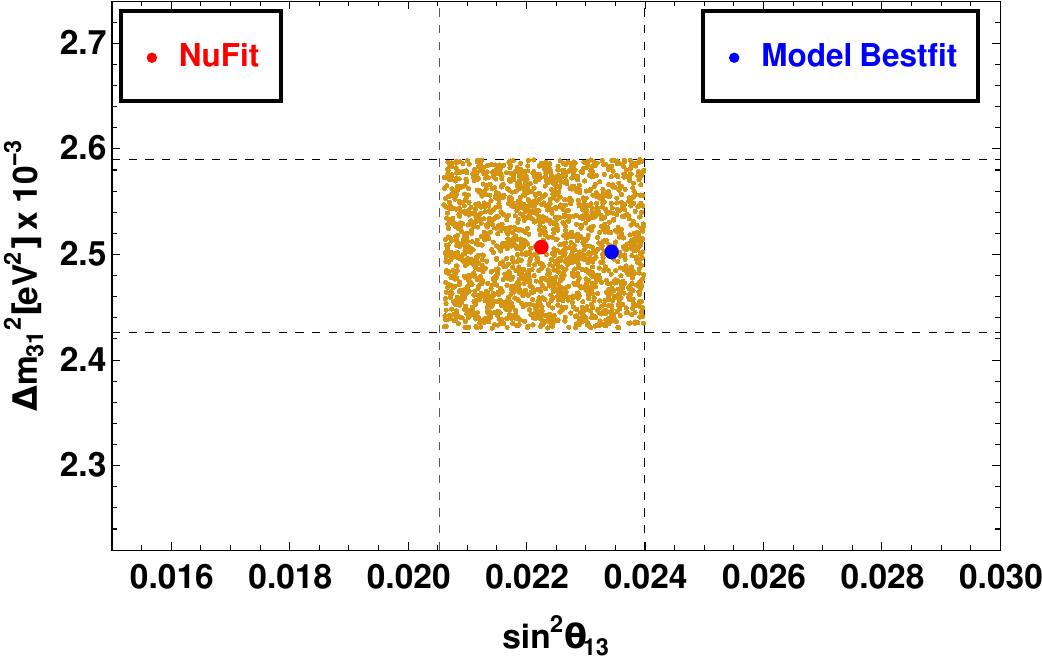}\label{ss13_dm13_model_A}}\\
    \subfloat[]{\includegraphics[height=45mm, width=54mm]{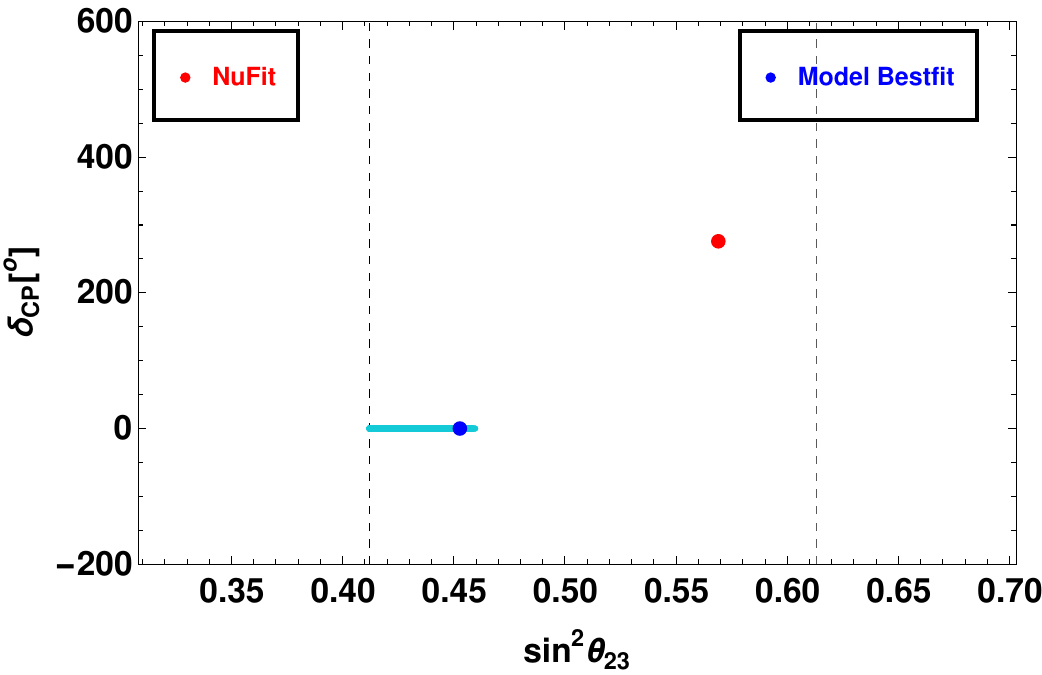}\label{ss23_dcp_model_B}}
    \subfloat[]{\includegraphics[height=45mm, width=54mm]{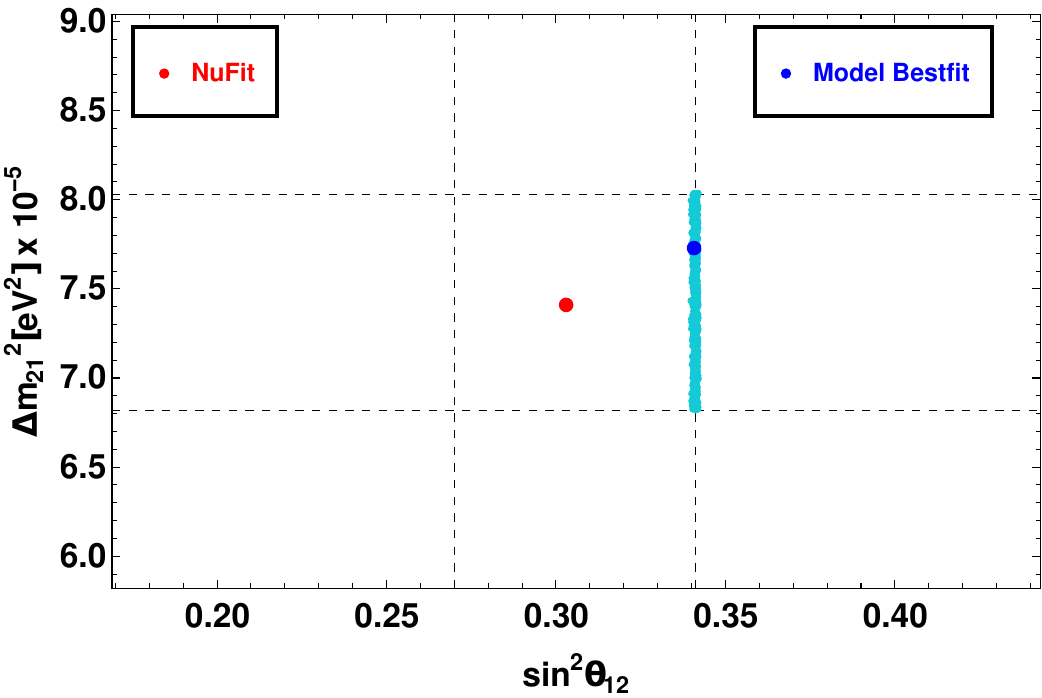}\label{ss12_dm12_model_B}}
    \hspace{0.1 cm}
    \subfloat[]{\includegraphics[height=45mm, width=54mm]{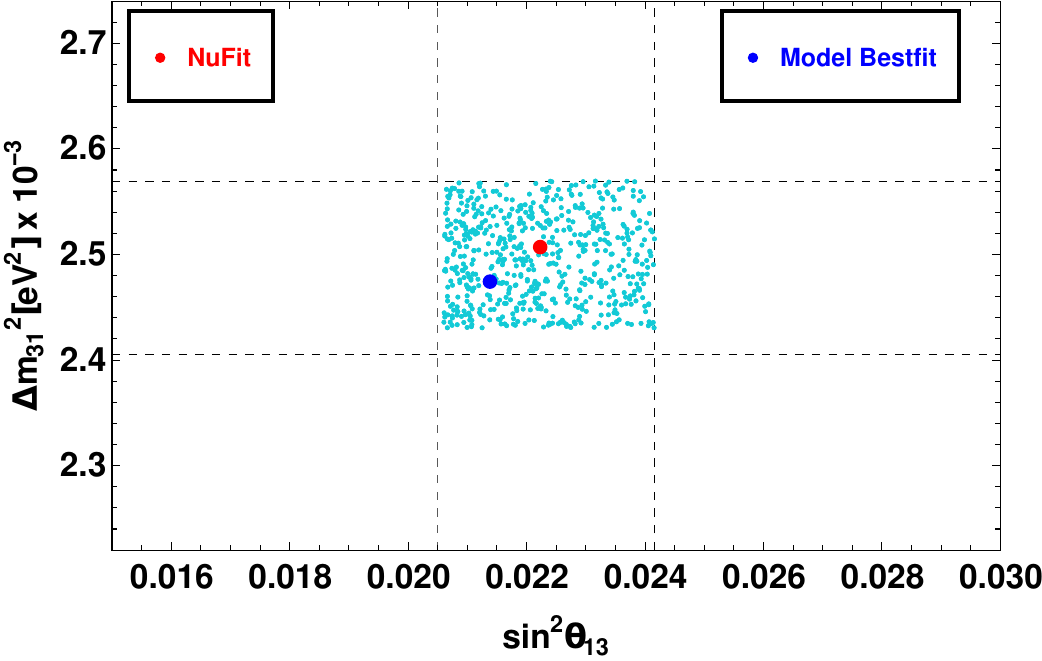}\label{ss13_dm13_model_B}}\\
    \subfloat[]{\includegraphics[height=45mm, width=54mm]{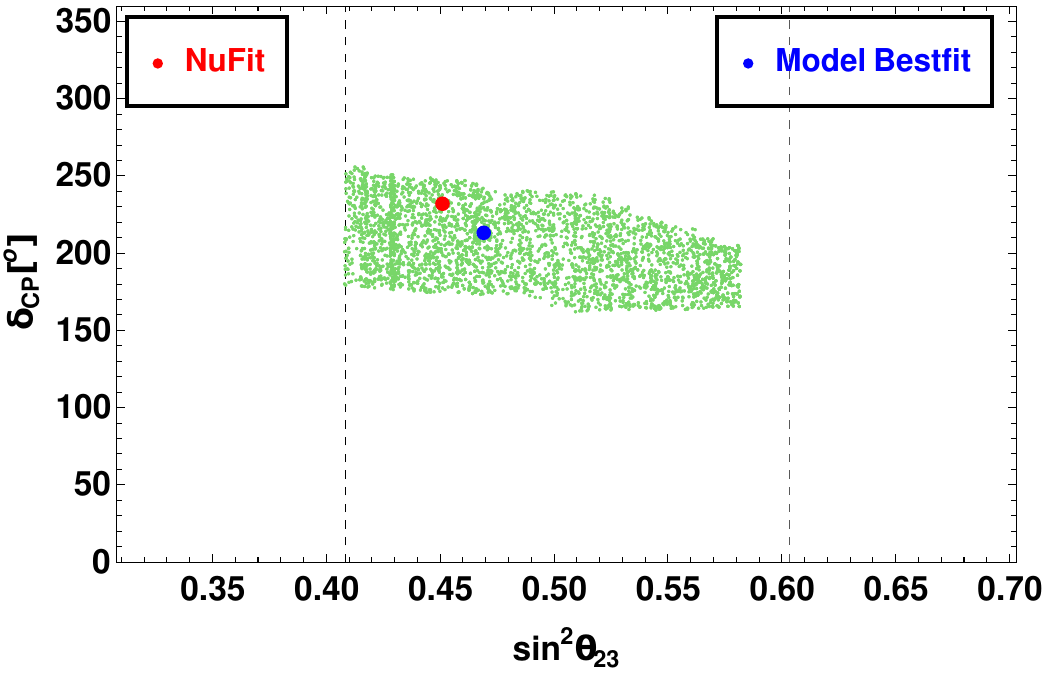}\label{ss23_dcp_model_C}}
    \subfloat[]{\includegraphics[height=45mm, width=54mm]{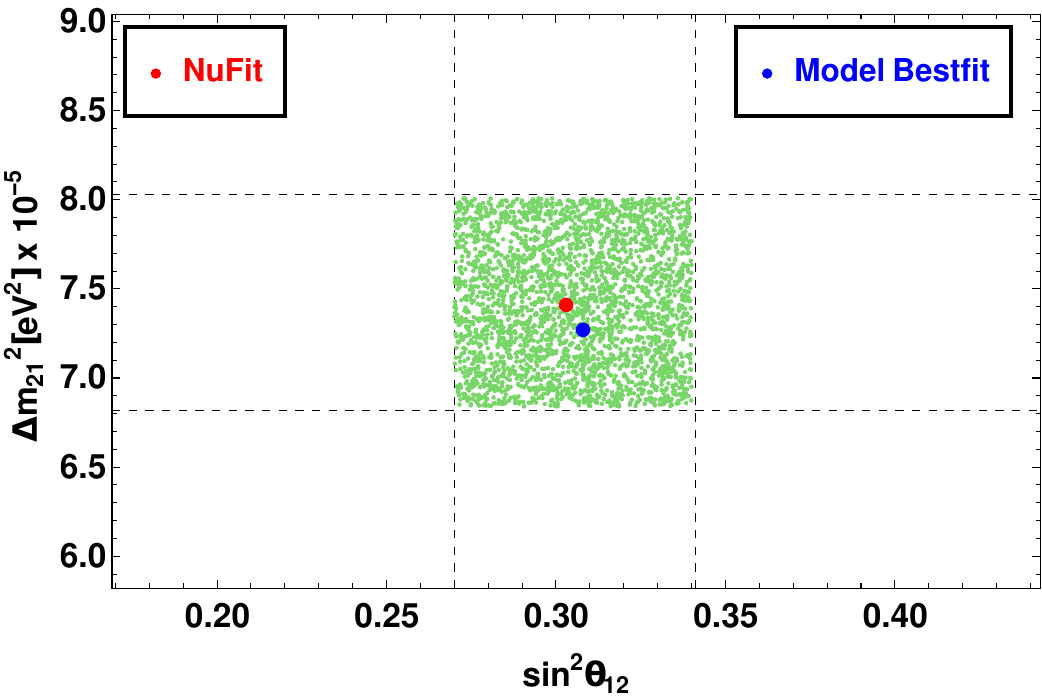}\label{ss12_dm12_model_C}}
    \hspace{0.1 cm}
    \subfloat[]{\includegraphics[height=45mm, width=54mm]{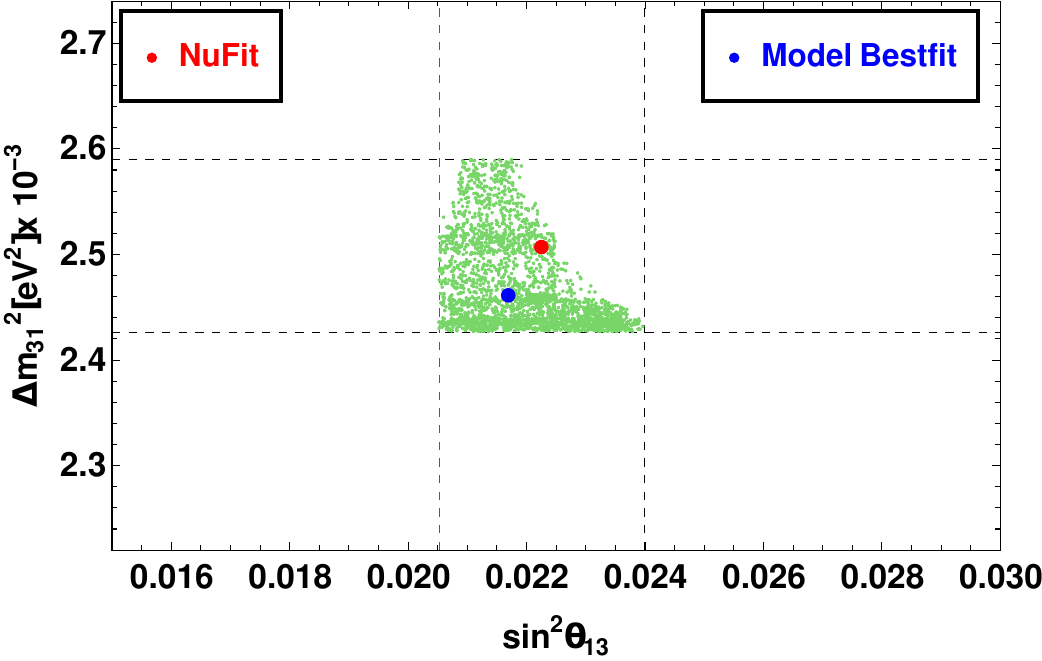}\label{ss13_dm13_model_C}}  
    \caption{The panels \ref{ss23_dcp_model_A}, \ref{ss23_dcp_model_B} and \ref{ss23_dcp_model_C} illustrate the correlation of $\sin^2 \theta_{23}$ w.r.t $\delta_{\rm CP}$ phase  for Model-A, B and C, respectively and \ref{ss12_dm12_model_A}, \ref{ss12_dm12_model_B} and \ref{ss12_dm12_model_C}  (\ref{ss13_dm13_model_A}, \ref{ss13_dm13_model_B} and \ref{ss13_dm13_model_C}) project the inter-dependence of $\Delta m_{21}^2$ ($\Delta m_{31}^2$) w.r.t to $\sin^2 \theta_{12}$ ($\sin^2 \theta_{13}$) for  Model-A, B and C respectively. Here the vertical and horizontal gridlines show the $3\sigma$ range of the corresponding oscillation parameters. Also, the NuFIT and model best-fit is shown by a red and blue dot, respectively.}
    \label{fig1}
\end{figure}
The parameter prediction of Model-C is shown in the bottom row of Fig.~\ref{fig1}. From the panels, we see that all the parameters except $\delta_{\rm CP}$ are unconstrained in this model. However, there exists a correlation between $\theta_{13}$ and $\Delta m^2_{31}$ as certain combinations of these parameters are not allowed. The allowed values of $\delta_{\rm CP}$ for this model lies in the range of $162^\circ < \delta_{\rm CP} < 256^\circ$. As the $\delta_{\rm CP}$ values allowed from this model are very disjoint from the other two models, this model can be easily differentiated from them. The best-fit values of the oscillation parameters as obtained from this model are listed in the fifth column of Table~\ref{tab:Nufit}.  

\section{Experimental characteristics}
\label{sec:exp_char}

 To simulate the experiments T2HK, DUNE and JUNO, we use the software GLoBES \cite{Huber:2004ka,Huber:2007ji}. Below we describe the specification of these experiments that we use in our calculation. \\
 
DUNE (Deep Underground Neutrino Experiment) is a future accelerator-based long-baseline experiment that features a 1284.9 km baseline with the line-averaged Earth-matter constant density 2.84 g/cm$^3$ and a 40 kton liquid Argon time projection chamber as   far detector. The neutrino source for this experiment can provide $1.1 \times 10^{21}$ protons on target (POT) per year with a beam power of 1.2 MW. In this case, the flux will be an on-axis wide-band flux. Our estimation of the physics capabilities of this set-up is based on assuming 5 years of neutrino run and 5 years of anti-neutrino run.  The specifications regarding backgrounds, systematic errors, etc., are taken from \cite{DUNE:2021cuw}.\\

T2HK (Tokai-to-Hyper-Kamiokande) is another upcoming long baseline experiment with baseline length of 295 km and is off-axial by  $2.5^\circ$, producing a  very narrow beam. For T2HK, we have used the configuration provided in Ref.~\cite{Hyper-Kamiokande:2016srs}. The neutrino source, stationed at J-PARC, will operate at a beam power of 1.3 MW and will have a total exposure of $27 \times 10^{21}$ protons on target (POT), equivalent to ten years of operation. The 10 years of runtime has been divided into two equal periods of five years each, one for neutrino and the other for antineutrino mode.   
The averaged Earth matter density for this baseline is around $2.70$ g/cm$^3$, and the detector technology will be water Cherenkov having a fiducial volume of 374 kt.\\

 For JUNO, we follow the configuration as given in Ref.~\cite{JUNO:2015zny}. We consider a liquid scintillator detector having 20 kton fiducial mass located at a distance of around  53 km from Yangjiang and Taishan nuclear power plants. We have considered the energy resolution of $3\%/\sqrt{E~ (\rm MeV)}$. In this analysis, we consider all the reactor cores located at the same distance from the detector. We consider the run-time to be 6 years.
\section{Results}
\label{sec:results_simulation}

 In this section, we will present our results. First, we study the capability of T2HK, DUNE, and JUNO to constrain the three models that we discussed above with respect to the current data. Then we study the capability of these experiments to differentiate one model from another.

We will present our results in terms of $\chi^2$ analysis. We define the Poisson $\chi^2$ as:

\begin{align}
    \chi^2 = 2 \sum_j \left [ N^{\mathrm{th}}_j - N^{\mathrm{true}}_j - N^{\mathrm{true}}_j ~\mathrm{ln}\left (\frac{N^{\mathrm{th}}_j}{N^{\mathrm{true}}_j}\right)\right ],
\end{align}
where $j$ is the number of energy bins,  $N^{\mathrm{th}}_j$ is the number of events in the test spectrum and $N^{\mathrm{true}}_j$ is the number of events in the true spectrum.

 To test the models against the current values of oscillation parameters, we take the current values of the oscillation parameters in the true and the values of the oscillation parameters predicted from the models in the test. For each set of true parameters, we minimize the $\chi^2$ with respect to all sets of predicted parameters. We find the minimum $\chi^2$ i.e., $\chi^2_{min}$ for all sets of true parameters and calculate $\Delta \chi^2$ as $\chi^2 - \chi^2_{min}$. As, Model-A and Model-C are allowed for the normal ordering of the neutrino masses, for these models, we use the true values from NuFIT corresponding to normal ordering in the neutrino masses. However, for Model-B, we use the NuFIT values corresponding to the inverted ordering of the neutrino masses. The best-fit values from NuFIT for both normal and inverted ordering of the neutrino masses are listed in the first and second row of Table~\ref{tab:Nufit}.  

 \begin{figure}[htbp]
        \centering
       \vspace{-2.5cm}
\subfloat[]{\includegraphics[height=5.2cm, width=6.5cm]{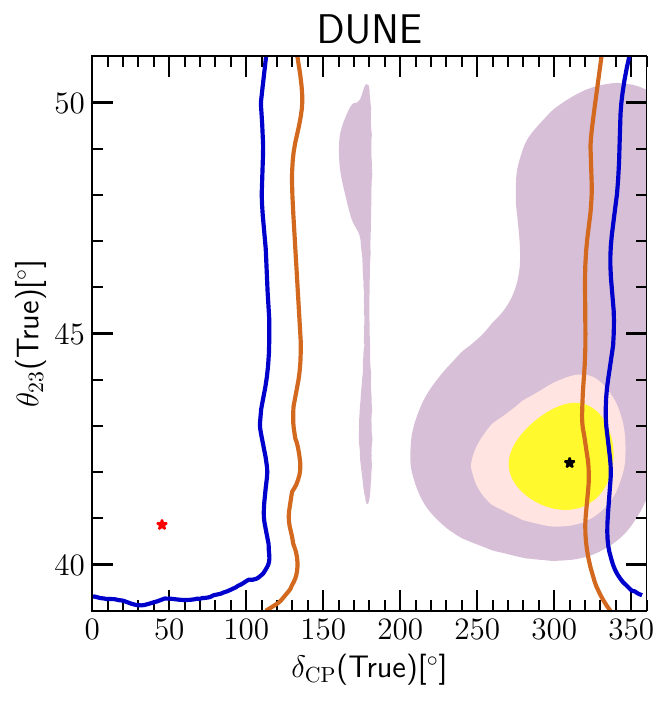}\label{DUNE-M1}}
\subfloat[]{\includegraphics[height=5.2cm, width=6.5cm]{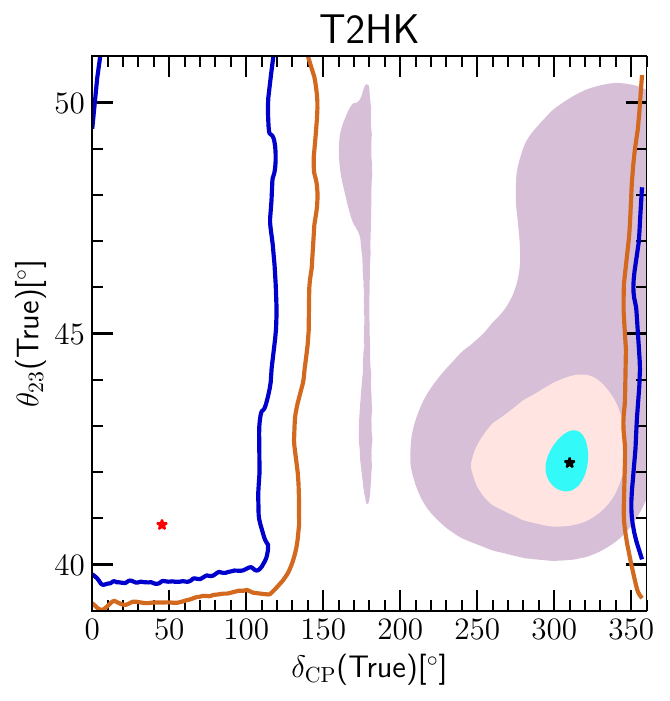}\label{T2HK-M1}}
\\ \subfloat[]{\includegraphics[height=5.2cm, width=6.5cm]{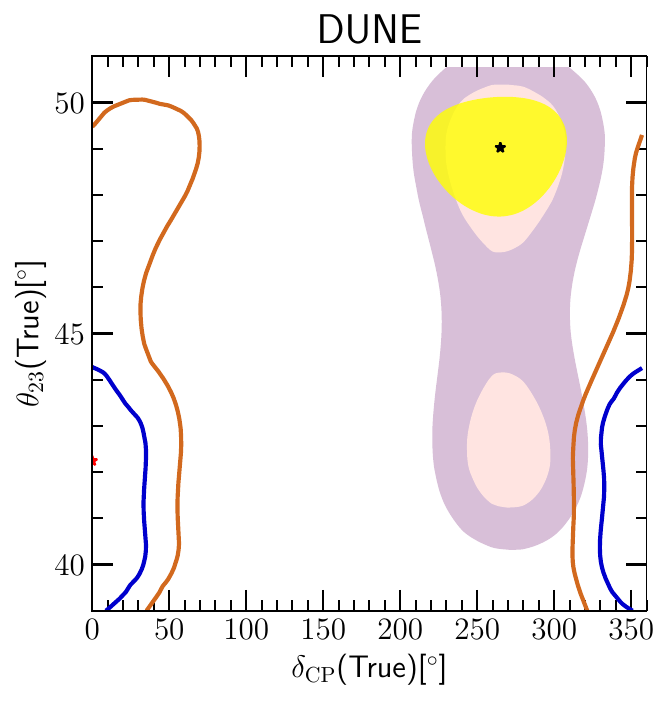}\label{DUNE-M2}}
\subfloat[]{\includegraphics[height=5.2cm, width=6.5cm]{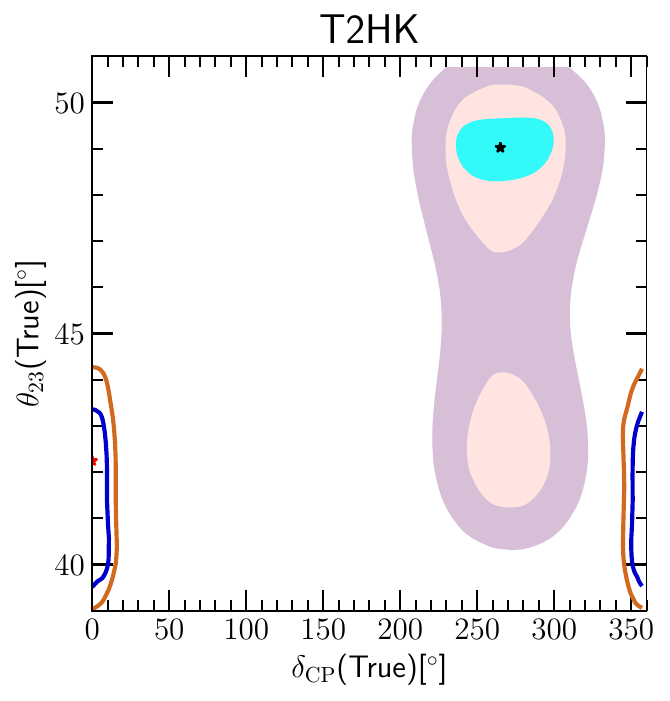}\label{T2HK-M2}}
\\ \subfloat[]{\includegraphics[height=5.2cm, width=6.5cm]{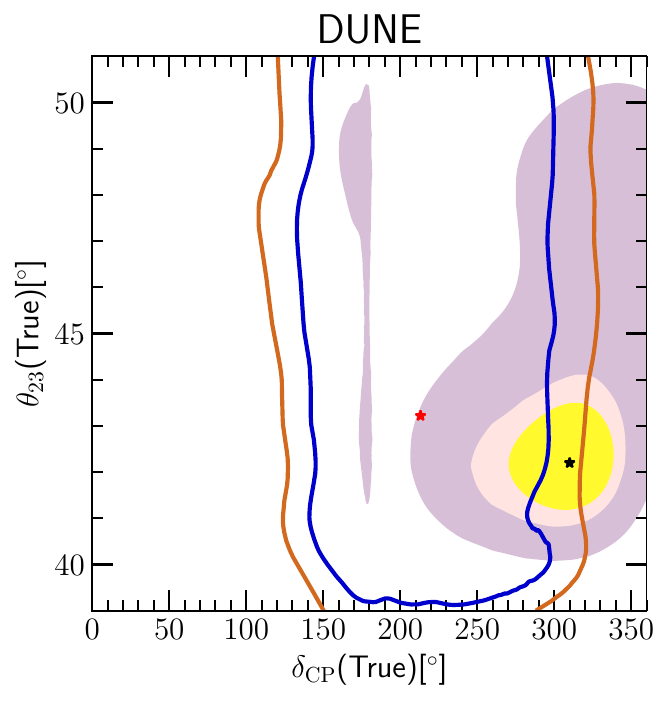}\label{DUNE-M3}}
\subfloat[]{\includegraphics[height=5.2cm, width=6.5cm]{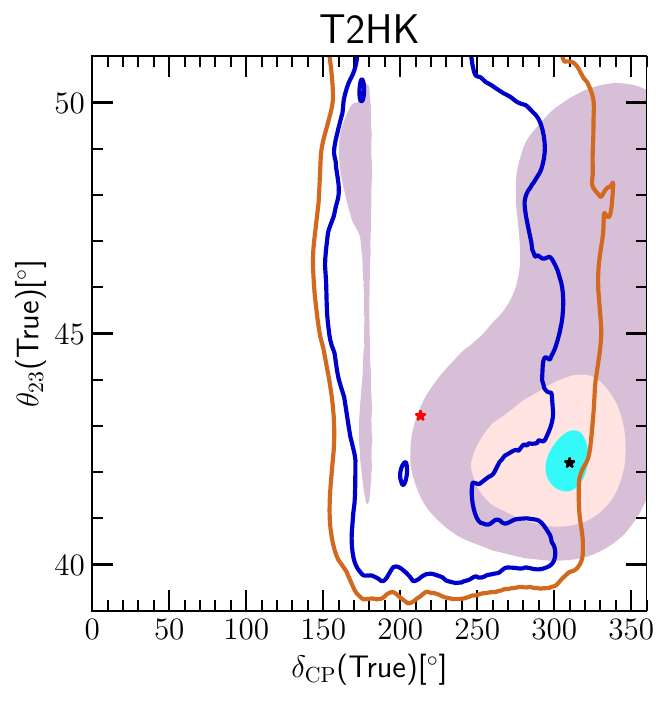}\label{T2HK-M3}} \\ \includegraphics[height=4.cm, width=5.5cm]{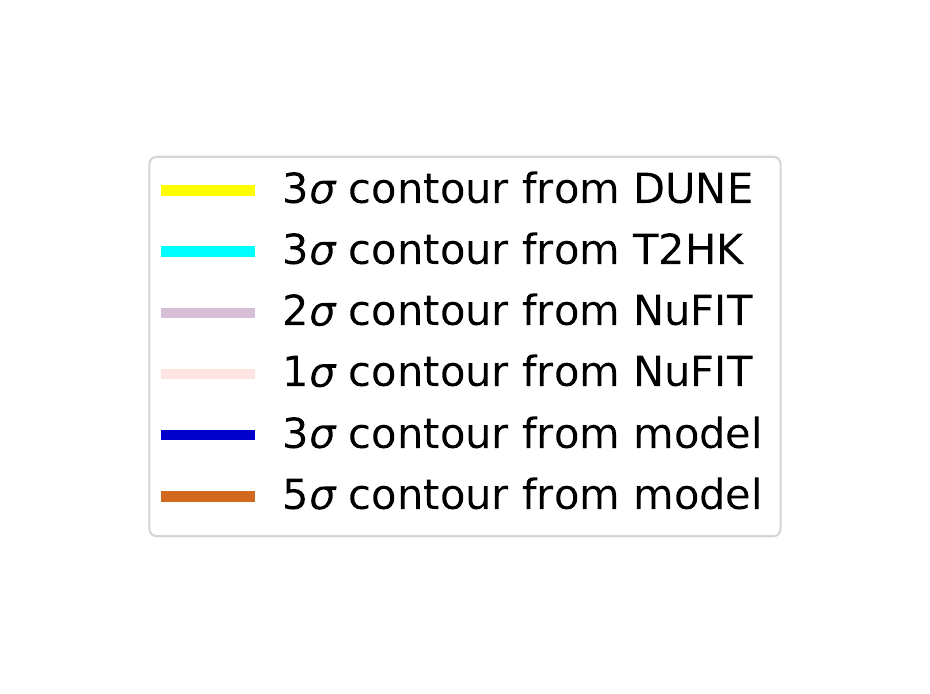} \\
\caption{Panels representing the sensitivity of DUNE and T2HK in $\theta_{23}$ (true) - $\delta_{\rm CP}$ (true) plane. Panel \ref{DUNE-M1}, \ref{DUNE-M2} and \ref{DUNE-M3} represent the compatibility of Model-A, B, and C with DUNE, respectively, while panel \ref{T2HK-M1}, \ref{T2HK-M2} and \ref{T2HK-M3} represent the compatibility of the same models with T2HK. The black and red asterisks correspond to the NuFIT and model best-fit, respectively.}
\label{fig:2}
\end{figure}

 These results are presented in Fig.~\ref{fig:2}. The first row is for Model-A, the second row is for Model-B, and the third row is for Model-C. For all three models, we have presented the sensitivity in the $\theta_{23}$ (true) - $\delta_{\rm CP}$ (true) plane for DUNE and T2HK. In those panels, we have shown the allowed regions corresponding to $3 \sigma$ (blue curve) and $5 \sigma$ C.L (brown curve). The best-fit values of $\theta_{23}$ and $\delta_{\rm CP}$ for these allowed regions are indicated by a red asterisk.  Additionally, in these panels, the current $1 \sigma$ and $2 \sigma$ allowed regions from NuFIT are shown in the orange and purple-shaded regions, respectively. The best-fit values from the NuFIT is shown by the black asterisk. Assuming the current best-fit values of $\theta_{23}$ and $\delta_{\rm CP}$ remain the same in future, DUNE and T2HK will measure these values very precisely. The yellow shaded and the cyan shaded region show the allowed region from DUNE and T2HK at $3 \sigma$ C.L, respectively, with respect to the current best-fit value of $\theta_{23}$ and $\delta_{\rm CP}$. Therefore, the yellow and cyan regions show how much the orange and the purple region will shrink in the future by the measurements of DUNE and T2HK. In the next paragraph, we will discuss the capability of T2HK and DUNE to exclude these models, assuming the current best-fit values remain the same in the future and they are measured by T2HK and DUNE. Or in other words, we will compare the allowed regions given by the blue and brown curves with respect to the yellow region for DUNE and the cyan region for T2HK.

From this figure, in general, we see that the allowed regions for all three models are more constrained in T2HK as compared to DUNE. This is probably because of the large statistics of T2HK, which arises due to its shorter baseline as compared to DUNE. For Model-A, we see that the $5 \sigma$ allowed region is well separated from the $3 \sigma$ allowed region for T2HK, but for DUNE, they are consistent. However, the $3 \sigma$ C.L. region of Model-A will not be compatible with the  $3 \sigma$ allowed region of DUNE. For Model-B, the $5 \sigma$ allowed region will not overlap with the $3 \sigma$ allowed region of both T2HK and DUNE if the future best-fit values of $\theta_{23}$ and $\delta_{\rm CP}$ remain same as the current best-fit value. Further, it will be difficult for both T2HK and DUNE to exclude Model-C as, in this case, the $5 \sigma$ allowed region of this model remains consistent with the $3 \sigma$ C.L. allowed regions of T2HK and DUNE. However, the $3 \sigma$ allowed region of Model-C is incompatible with the $3 \sigma$ allowed region of T2HK.

\begin{figure}[htbp]
        \centering
{\includegraphics[height=6.5cm, width=7.5cm]{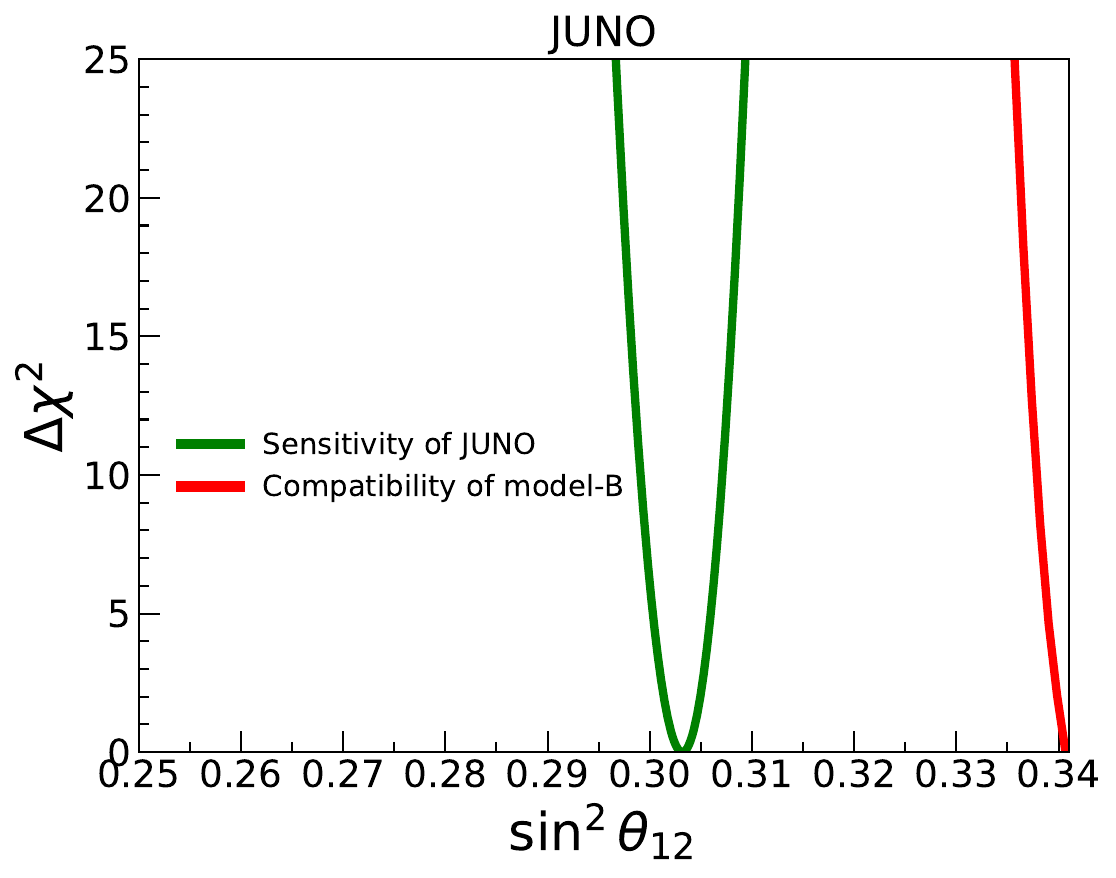}}
\caption{Figure illustrating the sensitivity of JUNO to constrain Model-B. The green solid curve shows the future precision of $\theta_{12}$ with respect to the current best fit of this parameter. The red solid curve represents allowed values of $\theta_{12}$ by JUNO for Model-B.} 
\label{fig:3}
\end{figure}

In the previous section, we have seen that Model-B predicts a very narrow range of $\theta_{12}$, and this gives us an opportunity to constrain this model from the future data of JUNO. In Fig.~\ref{fig:3}, we have shown the capability of JUNO to constrain this model (red curve). In this figure, the green curve shows the sensitivity of JUNO to constrain the parameter $\theta_{12}$, given its best-fit remains at the current best-fit value in the future.  From this figure, we observe that the future allowed values of $\theta_{12}$ by JUNO are very much separated from the allowed values of $\theta_{12}$ by Model-B. Therefore, JUNO alone will be able to exclude this model at an extremely high confidence level.

    \begin{figure}[htbp]
    \centering
    \includegraphics[height=55mm, width=70mm]{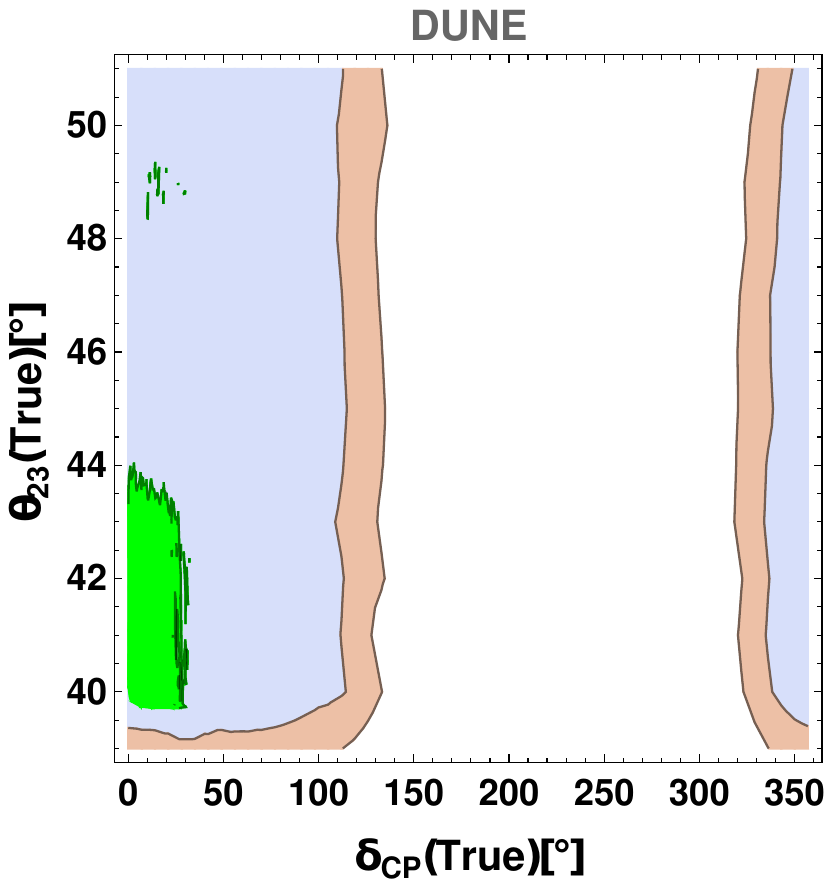}\label{M1-2_compare_dune_1}
   \includegraphics[height=55mm, width=70mm]{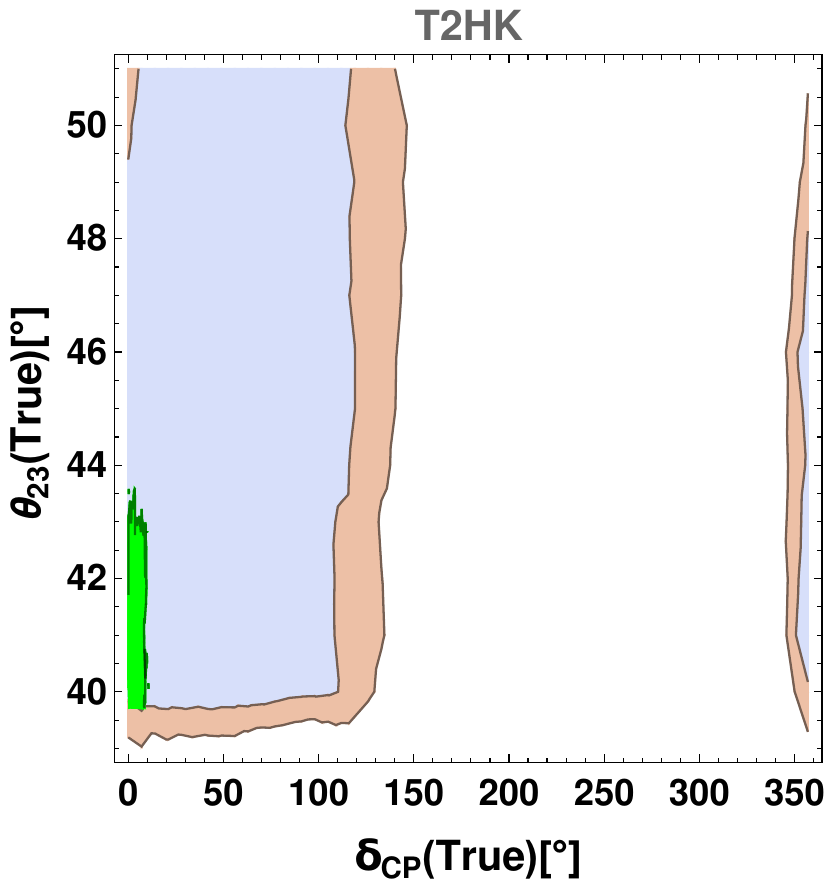}\label{M1-2_compare-T2HK-1.pdf}\\
     \label{comp-M1-2-V2}
    \caption{The blue (brown) shaded regions in the panels represent the $3\sigma$ ($5\sigma$) allowed regions of Model-A, while the green region corresponds to the parameter space for which Model-B cannot be separated from Model-A at  $3\sigma$ C.L. The left panel is for DUNE and the right panel is for T2HK.}
        \label{fig:4}
    \end{figure}
 Now let us discuss the capability of DUNE and T2HK to separate one model from another. In the previous section, we have seen that Model-A and Model-B have some common allowed parameter space in terms of $\theta_{23}$ and $\delta_{\rm CP}$. However, the allowed region of Model-C in $\theta_{23}$ - $\delta_{\rm CP}$ parameter space is disjoint with respect to Model-A and Model-B. This implies that Model-C can be easily separated from Model-A and Model-B in DUNE and T2HK, whereas these experiments will have confusion to separate Model-A from Model-B. We have shown this in Fig.~\ref{fig:4}. The left panel is for DUNE, and the right panel is for T2HK. To generate this figure, we have taken Model-A in the true spectrum of the $\chi^2$ and Model-B in the test spectrum. For each set of true parameters for Model-A, we have minimized over all sets of test parameters from Model-B and presented the results in the $\delta_{\rm CP}$ (true) - $\theta_{23}$ (true) plane for Model-A. Here also, we have subtracted the overall $\chi^2_{min}$ from each $\chi^2$ to calculate the $\Delta \chi^2$. This $\Delta \chi^2$ is represented in these panels as green allowed regions at $3 \sigma$ C.L.  In these panels, blue and brown regions are the allowed region for Model-A at $3 \sigma$ C.L and $5 \sigma$ C.L., respectively. Therefore, we understand that in the green region, DUNE, and T2HK will not be capable of separating Model-A from Model-B. However, in the rest of the blue and brown regions, DUNE and T2HK will be able to distinguish between Model-A and Model-B. Here also, we see that the confusion between Model-A and Model-B will be more prominent in DUNE as compared to T2HK.
\section{Conclusion}
\label{sec:conclusion}
This paper explores the potential of upcoming neutrino experiments to constrain a series of theoretical models that utilize modular symmetry. Here, we make use of $A_4$ modular symmetry, which is favorable in eliminating the excess usage of flavon fields. In lieu of traditional flavon fields, the model incorporated modular Yukawa couplings that transformed non-trivially under a modular $A_4$ group, resulting in a unique flavor structure of the neutrino mass matrix and facilitating the study of neutrino phenomenology. The models under scrutiny are linear seesaw identified as Model-A and involves three right-handed (RH) ($N_{R}$) and three left-handed (LH) sterile ($S_{L}$) superfields and a weighton ($\rho_a$); Model-B is a type-I seesaw utilizing three RH ($N_{R}$) without any flavons, and Model-C is a type-III seesaw involving fermion triplets superfields $(\Sigma_{R})$ and a weighton ($\rho_c$) in supersymmetric context. These models based on modular symmetry can predict the values of the leptonic mixing parameters, and therefore, there is an opportunity to probe these models in the neutrino oscillation experiments. 

We choose our models in such a way that their predictions of the neutrino oscillation parameters are different from each other, but these models are allowed within the currently allowed ranges of these parameters. Among the three models, Model-A and Model-C are allowed for the normal ordering of the neutrino masses, whereas Model-B is allowed for the inverted ordering of the neutrino masses. For our analysis, we have considered the future accelerator-based neutrino experiments T2HK and DUNE and also the medium-baseline reactor experiment JUNO. The experiments T2HK and DUNE are expected to measure the parameters $\theta_{23}$ and $\delta_{\rm CP}$ with excellent precision, whereas  JUNO will provide the most stringent measurement of $\theta_{12}$. As the three models that we consider in our study provide narrow ranges of the oscillation parameters $\theta_{23}$, $\theta_{12}$ and $\delta_{\rm CP}$, the experiments T2HK, DUNE and JUNO will have the capability to constrain these models in future. Apart from that, it is also expected that DUNE, T2HK and JUNO will discover the true nature of the neutrino mass ordering.  Note that once the true mass ordering of the neutrino masses is determined, either Model-A and Model-C or Model-B can be excluded immediately. However, in this work, we will assume that neutrino mass ordering will be unknown, and we will study these models in future neutrino experiments based on their prediction of the neutrino oscillation parameters. For our models, it is not possible to define the sum rules which relate the model parameters to the leptonic mixing parameters. Therefore we adopted the strategy to estimate the prediction of the models numerically and then used them as input for studying these models in the neutrino experiments.

Among the three models that we consider in our study, Model-A predicts the values of $\delta_{\rm CP}$ to lie in the range of $0^\circ \le \delta_{\rm CP} < 89^\circ$. Model-B allows only $\delta_{\rm CP} = 0^\circ$ and predicts lower octant of $\theta_{23}$ in the region $\sin^2\theta_{23} < 0.459$. It also predicts a narrow range of $\theta_{12}$ with $\sin^2\theta_{12}$ around 0.34. The allowed values of $\delta_{\rm CP}$ for Model-C lie in the range of $162^\circ < \delta_{\rm CP} < 256^\circ$. When studying these models in the neutrino experiments, we find that capability of T2HK to constrain the models is better as compared to DUNE. Assuming the future best-fit of $\theta_{23}$ and $\delta_{\rm CP}$ remains the same as the current one, we noticed that T2HK could exclude Model-A at more than $5 \sigma$ C.L, but this model will be consistent with DUNE at $5 \sigma$ C.L.  For Model-B, both T2HK, and DUNE will be capable of excluding this model at more than $5 \sigma$ C.L. Model-C cannot be excluded by T2HK and DUNE at $5 \sigma$ C.L. Further, our results show that JUNO alone can exclude Model-B at an extremely high confidence level if the future best-fit of $\theta_{12}$ remains at the current-one. As Model-A and Model-B have a common parameter space in terms of $\theta_{23}$ and $\delta_{\rm CP}$, we tried to see if these models can be distinguished by T2HK and DUNE. We have identified the region in the $\theta_{23}$ - $\delta_{\rm CP}$ parameter space for which Model-A cannot be separated from Model-B in T2HK and DUNE. 

In summary, our results demonstrated the capability of T2HK, DUNE, and JUNO to constrain a set of theoretical models based on modular symmetry for which it is not possible to have sum rules connecting the model parameters and the leptonic mixing parameters. These results will have a significant impact in the future in terms of excluding theoretical models in future experiments.

\section{Acknowledgement}

PM and PP want to thank Prime Minister's Research Fellows (PMRF) scheme for its financial support. MKB would like to acknowledge Program Management Unit for Human Resources \& Institutional Development, Research, and Innovation (PMU-B). This academic position is administered in the framework of the F13 (S4P21) National Postdoctoral/Postgraduate System, contract no. B13F660066. RM would like to acknowledge University of Hyderabad IoE project grant no. RC1-20-012. This work has been in part funded by the Ministry of Science and Education of the Republic of Croatia grant No. KK.01.1.1.01.0001. We gratefully acknowledge the use of the CMSD HPC facility of Univ. of Hyderabad to carry out the computational work.
\bibliographystyle{my-JHEP}
\bibliography{probe}

\end{document}